\documentclass[aps,prc,preprint,showpacs,showkeys]{revtex4}
\usepackage{amssymb}
\usepackage{amsmath}
\usepackage{graphicx}
\usepackage{bm}
\usepackage{epsf}
\usepackage{ulem}
\usepackage{color, soul}

\begin{document}

\title{Finite-size effects on the hadron-quark phase transition
       in neutron stars}
\author{X. H. Wu}
\affiliation{School of Physics, Nankai University, Tianjin 300071, China}
\author{H. Shen}~\email{shennankai@gmail.com}
\affiliation{School of Physics, Nankai University, Tianjin 300071, China}

\begin{abstract}
We study the finite-size effects, like the surface and Coulomb energies,
on the hadron-quark mixed phase in neutron stars.
The equilibrium conditions for coexisting hadronic and quark phases are
derived by minimizing the total energy including the surface and
Coulomb contributions, which are different from the Gibbs conditions
without finite-size effects.
We employ the relativistic mean-field model to describe the hadronic
phase, while the Nambu-Jona-Lasinio model with vector interactions is used
for the quark phase. It is found that finite-size effects can significantly
reduce the region of the mixed phase, and the results lie between those of
the Gibbs and Maxwell constructions. We show that a massive star may contain
a mixed phase core and its size depends on the surface tension of the
hadron-quark interface and the vector coupling between quarks.
The repulsive vector interaction in the Nambu-Jona-Lasinio model
can stiffen the equation of state of quark matter, and therefore, delay
the phase transition and increase the maximum mass of neutron stars.

\end{abstract}

\maketitle


\section{Introduction}
\label{sec:1}

Neutron stars provide a unique environment for the study of cold and
dense matter. It is expected that the phase transition from hadronic
matter to quark matter may occur in the core of massive neutron
stars~\cite{Glen01,Heis00,Webe05}.
Over the past decades, many authors have studied the deconfinement
phase transition of neutron-star matter and its influence on properties
of neutron stars~\cite{Glen92,Sche99,Sche00,Latt00,Burg02,Mene03,Shar07,
Yang08,Xu10,Chen13,Orsa14}. Most of these studies are based on the bulk
approximation, in which the hadron-quark phase transition is performed
through the Maxwell or Gibbs constructions.
In the Maxwell construction, local charge neutrality is imposed,
while the coexisting hadronic and quark phases have equal pressure and
baryon chemical potential but different electron chemical potential.
The pressure of the mixed phase in the Maxwell construction remains constant,
and therefore such a mixed phase is not allowed to appear inside neutron stars.
With the Gibbs conditions for phase equilibrium, only global charge
neutrality is required, while hadronic and quark phases have opposite
electric charges. The mixed phase in the Gibbs construction persists
over a finite range of pressure, so it is possible for the massive
neutron star to contain a mixed-phase region in its interior.
It has been shown in Ref.~\cite{Bhat10} that there are significant
differences in the behavior of compact stars between Maxwell
and Gibbs constructions.

In general, the mixed phase with the Gibbs construction
is energetically more favorable than the one with the Maxwell construction.
However, when the surface and Coulomb energies are taken into account,
the energy density of the quark-droplet phase may
be higher than that of the Maxwell construction due to a large surface
tension~\cite{Heis93}. A detailed calculation including charge screening
indicates that the mixed phase with a large surface tension behaves
like that of the Maxwell construction, while the one with a small
surface tension is close to the case of the Gibbs
construction~\cite{Endo06,Maru07,Yasu14}.
In fact, the Maxwell and Gibbs constructions correspond to
the two limits of infinite and zero surface tension, respectively.
It is interesting to examine the effect of a finite surface tension
on properties of the hadron-quark mixed phase.

In an earlier study performed by Heiselberg~\textit{et al}.~\cite{Heis93},
the Coulomb and surface effects were examined for the description of
the hadron-quark mixed phase. The geometrical structure and size of
the mixed phase could be determined by competition between surface and
Coulomb energies. The possible geometrical structure of the mixed phase
has been extensively discussed
in Refs.~\cite{Endo06,Maru07,Yasu14,Glen95,Glen97},
which may change from droplet to rod, slab, tube, and bubble
with increasing density.
It was reported in Refs.~\cite{Endo06,Vosk05} that the charge screening effect
and the rearrangement of charged particles should be taken into account
for a realistic description of the hadron-quark mixed phase.
In most of the studies on the finite-size effects,
the coexisting hadronic and quark phases were required to
satisfy the Gibbs conditions for phase equilibrium.
An additional pressure due to the surface tension was also included in
the pressure equilibrium condition~\cite{Endo06}.
In fact, the equilibrium conditions would be modified when the
surface and Coulomb energies are taken into account.
Proper equilibrium conditions could be derived by minimizing the total free
energy of a system~\cite{Latt91,Bao14b}.
It is important to examine the equilibrium conditions between the hadronic and quark
phases with inclusion of the surface and Coulomb terms.

In the present work, we employ the Wigner--Seitz approximation to
describe the hadron-quark mixed phase. We derive the equilibrium conditions
for coexisting hadronic and quark phases by minimization
of the total energy including the surface and Coulomb contributions.
In the Wigner--Seitz cell, the hadronic and quark phases are assumed to be separated
by a sharp interface with a finite surface tension.
The surface tension plays a crucial role in determining the structure of
the mixed phase, but its value is poorly known.
The calculation in the MIT bag model by using the multiple reflection
expansion method~\cite{Berg87} gave a value of the surface tension
$\sigma \sim 10$ MeV/fm$^2$, while a similar calculation in the
Nambu-Jona-Lasinio (NJL) model including color superconductivity~\cite{Lugo13}
yielded $\sigma \sim 145$--165 MeV/fm$^2$.
The surface tension calculated from a geometrical approach fell in
the range $\sigma \sim 7$--30 MeV/fm$^2$~\cite{Pint12}.
Considering the uncertainty of $\sigma$, we treat the surface tension
as a free parameter in the present study following the idea
of Refs.~\cite{Heis93,Endo06,Yasu14}. Furthermore, the finite-size effects
on the hadron-quark phase transition are examined by varying this parameter.
The limit of $\sigma=0$ corresponds to the absence of surface and Coulomb
energies in the Gibbs construction.

The aim of the present work is to investigate the influence of
surface and Coulomb contributions on the hadron-quark phase transition
using the equilibrium conditions derived by minimization of the total energy.
We employ the relativistic mean-field (RMF) model to describe the
hadronic phase, while the NJL model is used for the quark phase.
The NJL model can successfully describe dynamical chiral symmetry breaking
and generation of constituent quark masses, so it has been widely used as an
effective theory of QCD for the description of quark
matter~\cite{Mene03,Hats94,Buba05,Logo13,Bona12}.
In this work, we adopt the three-flavor NJL model with a repulsive
vector interaction. It has been extensively discussed in the literature
that the inclusion of repulsive vector interactions could significantly affect
the QCD phase diagram~\cite{Fuku08,Ueda13,Buba15} and stiffen the equation of
state (EOS) of quark matter which would result in larger maximum neutron-star
masses~\cite{Yasu14,Logo13,Bona12,Pagl08,Abuk09,Masu13,Mene14,Hell14,Chu15,Pere16}.
For hadronic matter, we employ the RMF model with the parameter
set TM1~\cite{Suga94}, which can satisfactorily describe the properties of
nuclear matter and finite nuclei. This model has been successfully applied to
construct the EOS for supernova simulations and
neutron stars~\cite{Shen02,Shen11}. With only nucleonic degrees of freedom,
the TM1 model predicts a maximum neutron-star mass of $2.18\ M_{\odot}$.
If $\Lambda$ hyperons are allowed to appear, the maximum mass is reduced
to $1.75\ M_{\odot}$~\cite{Shen11}.
It is well known that the appearance of hyperons
can significantly soften the EOS at high density and thus
reduce the maximum neutron-star mass.
The accurate mass determinations for PSR J1614-2230~\cite{Demo10,Fons16}
and PSR J0348+0432~\cite{Anto13} provide a strong constraint on
the EOS of neutron-star matter.
Most of the EOS including hyperons cannot satisfy the maximum mass
constraint. It has been reported in Ref.~\cite{Masu13} that
by using the EOS interpolated between hadronic matter with hyperons
and quark matter in a crossover region, the maximum neutron-star mass
could be compatible with the observations, and an earlier onset of the
hadron-quark crossover would provide a larger maximum mass.
The authors of Ref.~\cite{Masu13} considered several different
hadronic EOSs obtained by $G$-matrix calculations and the chiral SU(3)
symmetric RMF model, and they found that the qualitative conclusion
is insensitive to the choice of the hadronic EOS.
Currently there are large uncertainties in the contributions
from hyperons at high density~\cite{Weis12}.
Therefore, we do not include hyperons in the present calculation
and focus on the transition from nonstrange hadronic matter
to deconfined quark matter with the inclusion of finite-size effects.

This article is organized as follows.
In Sec.~\ref{sec:2}, we briefly describe the RMF model for hadronic matter.
In Sec.~\ref{sec:3}, the NJL model used for quark matter is shortly introduced.
In Sec.~\ref{sec:4}, we describe the hadron-quark mixed phase with
finite-size effects and derive the equilibrium conditions for coexisting
phases by minimization of the total energy including the surface and Coulomb
contributions.
In Sec.~\ref{sec:5}, we present the numerical results
and discuss the finite-size effects on the hadron-quark phase transition
and neutron star properties.
Section~\ref{sec:6} is devoted to the conclusions.

\section{Hadronic matter phase}
\label{sec:2}

We employ the RMF model to describe the hadronic
matter phase. In the RMF approach, nucleons interact via the exchange of
various mesons. The exchanged mesons considered here include the
isoscalar scalar and vector mesons ($\sigma$ and $\omega$)
and isovector vector meson $\rho$.
We adopt the RMF model with the parameter set TM1, which provides an
excellent description of nuclear matter and finite nuclei.
For hadronic matter consisting of nucleons ($p$ and $n$) and
leptons ($e$ and $\mu$), the effective Lagrangian reads
\begin{eqnarray}
\label{eq:LRMF}
\mathcal{L}_{\rm{RMF}} & = & \sum_{i=p,n}\bar{\psi}_i
\left( i\gamma_{\mu}\partial^{\mu}- M - g_{\sigma}\sigma
  -g_{\omega}\gamma_{\mu}\omega^{\mu}
  -g_{\rho}\gamma_{\mu}\tau_a\rho^{a\mu}\right)\psi_i  \notag \\
&& +\frac{1}{2}\partial_{\mu}\sigma\partial^{\mu}\sigma
  -\frac{1}{2}m^2_{\sigma}\sigma^2
  -\frac{1}{3}g_{2}\sigma^{3} -\frac{1}{4}g_{3}\sigma^{4} \notag \\
&& -\frac{1}{4}W_{\mu\nu}W^{\mu\nu} +\frac{1}{2}m^2_{\omega}\omega_{\mu}
  \omega^{\mu} +\frac{1}{4}c_{3}\left(\omega_{\mu}\omega^{\mu}\right)^2 \notag\\
&& -\frac{1}{4}R^a_{\mu\nu}R^{a\mu\nu} +\frac{1}{2}m^2_{\rho}\rho^a_{\mu}
  \rho^{a\mu} \notag \\
&& +\sum_{l=e,\mu}\bar{\psi}_{l}
  \left( i\gamma_{\mu }\partial^{\mu }-m_{l}\right)\psi_l,
\end{eqnarray}
where $W^{\mu\nu}$ and $R^{a\mu\nu}$ are the antisymmetric field
tensors for $\omega^{\mu}$ and $\rho^{a\mu}$, respectively.
In the RMF approach, we treat the meson fields as classical fields
and replace them by their expectation values.
The nonvanishing expectation values of meson fields in hadronic
matter are $\sigma =\left\langle \sigma \right\rangle$,
$\omega =\left\langle\omega^{0}\right\rangle$,
and $\rho =\left\langle \rho^{30} \right\rangle$.
The equations of motion for the meson fields in uniform matter
are given by
\begin{eqnarray}
 m_{\sigma }^{2}\sigma +g_{2}\sigma ^{2}+g_{3}\sigma
^{3} &=& -g_{\sigma }\left( n_{p}^{s}+n_{n}^{s}\right) ,
\label{eq:eqms} \\
 m_{\omega }^{2}\omega +c_{3}\omega^{3}
&=& g_{\omega}\left( n_{p}+n_{n}\right) ,
\label{eq:eqmw} \\
 m_{\rho }^{2}{\rho}
&=& g_{\rho }\left(n_{p}-n_{n}\right) ,
\label{eq:eqmr}
\end{eqnarray}%
where $n_i^s$ and $n_i$ denote the scalar and number densities
of species $i$, respectively.
With the parameter set TM1 listed in Table~\ref{tab:1},
these coupled equations are solved self-consistently, which yield that
the nuclear matter saturation density is $0.145$ fm$^{-3}$,
the binding energy per nucleon is $16.3$ MeV,
the symmetry energy is $36.9$ MeV, and the incompressibility is $281$ MeV.

For hadronic matter in $\beta$ equilibrium, the chemical potentials
satisfy the relations $\mu_{p}=\mu_{n}-\mu_{e}$ and $\mu_{\mu}=\mu_{e}$.
At zero temperature, the chemical potentials of leptons are expressed by
$\mu_{l}=\sqrt{{k_{F}^{l}}^{2}+m_{l}^{2}}$, while that of nucleons are given by
$\mu_i=\sqrt{{k_{F}^{i}}^{2}+{M^{\ast}}^2}+g_{\omega}\omega
+g_{\rho}\tau_{3}^{i}\rho$ with $M^{\ast}=M+g_{\sigma}\sigma$
being the effective nucleon mass.
The total energy density of hadronic matter can be written as
\begin{eqnarray}
\label{eq:e1}
\varepsilon_{\rm{HP}} &=&\sum_{i=p,n}\frac{1}{\pi^{2}}
\int_{0}^{k_{F}^{i}} \sqrt{k^2+{M^{\ast}}^2}\ k^{2}dk
+\frac{1}{2}m_{\sigma }^{2}\sigma ^{2}+\frac{1}{3}g_{2}\sigma ^{3}
+\frac{1}{4}g_{3}\sigma ^{4}  \nonumber \\
&&+\frac{1}{2}m_{\omega }^{2}\omega ^{2}+\frac{3}{4}c_{3}\omega ^{4}
+\frac{1}{2}m_{\rho }^{2}\rho^{2}
+\sum_{l=e,\mu}\frac{1}{\pi^{2}}\int_{0}^{k_{F}^{l}}
\sqrt{k^{2}+m_{l}^{2}}\ k^{2}dk,
\end{eqnarray}
and the pressure is given by
\begin{eqnarray}
\label{eq:p1}
P_{\rm{HP}} &=& \sum_{i=p,n}\frac{1}{3\pi^{2}}
\int_{0}^{k_{F}^{i}}\frac{k^{4} dk}{\sqrt{k^{2}+{M^{\ast}}^2}}
-\frac{1}{2}m_{\sigma }^{2}\sigma^{2}-\frac{1}{3}g_{2}\sigma ^{3}
-\frac{1}{4}g_{3}\sigma ^{4}  \nonumber \\
&&+\frac{1}{2}m_{\omega }^{2}\omega ^{2}+\frac{1}{4}c_{3}\omega^{4}
+\frac{1}{2}m_{\rho }^{2}\rho^{2}
+\sum_{l=e,\mu}\frac{1}{3\pi^{2}}\int_{0}^{k_{F}^{l}}
\frac{k^{4} dk}{\sqrt{k^{2}+m_{l}^{2}}}.
\end{eqnarray}

\section{Quark matter phase}
\label{sec:3}

For the description of quark matter, we employ the NJL
model with three flavors. The Lagrangian is written as
\begin{eqnarray}
\label{eq:Lnjl}
\mathcal{L}_{\rm{NJL}} &=&\bar{q}\left( i\gamma _{\mu }\partial ^{\mu
}-m^{0}\right) q+{G_S}\sum\limits_{a = 0}^8 {\left[ {{{\left( {\bar q{\lambda _a}q} \right)}^2}
+ {{\left( {\bar q i{\gamma _5}{\lambda _a}q} \right)}^2}} \right]}  \nonumber \\
&&-K\left\{ \det \left[ \bar{q}\left( 1+\gamma _{5}\right) q\right] +\det %
\left[ \bar{q}\left( 1-\gamma _{5}\right) q\right] \right\} \nonumber \\
&&- {G_V}\sum\limits_{a = 0}^8 {\left[ {{{\left( {\bar q{\gamma ^\mu }{\lambda _a}q} \right)}^2}
+ {{\left( {\bar q{\gamma ^\mu }{\gamma _5}{\lambda _a}q} \right)}^2}} \right]},
\end{eqnarray}%
where $q$ denotes the quark field with three flavors and three colors.
The first term is the free Dirac Lagrangian with the current quark
mass matrix given by $m^{0}=\text{diag} \left(m_{u}^{0},m_{d}^{0},m_{s}^{0}\right)$.
The second term with coupling ${G_S}$ is a chirally symmetric four-quark interaction,
where $\lambda_a$ are the flavor SU(3) Gell-Mann matrices with $\lambda_0=\sqrt{2/3}\,I$.
The third term corresponds to the six-quark Kobayashi--Maskawa--'t Hooft interaction that
breaks the $U_A(1)$ symmetry. The last term introduces additional vector and axial-vector
interactions with a positive coupling ${G_V}$ that play important roles in describing
massive stars~\cite{Abuk09,Masu13,Mene14,Hell14,Chu15,Pere16}.
In the present work, we adopt the parameters given in Ref.~\cite{Rehb96},
$m_{u}^{0}=m_{d}^{0}=5.5\ \text{MeV}$, $m_{s}^{0}=140.7\ \text{MeV}$,
$\Lambda =602.3\ \text{MeV}$, ${G_S}\Lambda^{2}=1.835$,
and $K\Lambda ^{5}=12.36$. As for the vector coupling ${G_V}$, we treat it as a free
parameter and take the ratios ${G_V}/{G_S}=0$, 0.2, and 0.4, in order to
investigate the effect of the repulsive vector interaction on the equation of state.

In the NJL model at the mean-field level, the quarks get constituent quark
masses by spontaneous chiral symmetry breaking. The constituent quark mass
in vacuum $m_{i}$ is considerably larger than the current quark mass $m_{i}^{0}$.
In quark matter, the constituent quark masses $m_{i}^{\ast }$ are determined
from the coupled set of gap equations
\begin{equation}
\label{eq:gap}
m_{i}^{\ast }=m_{i}^{0}-4{G_S}\langle \bar{q}_{i}q_{i}\rangle +2K\langle \bar{q}%
_{j}q_{j}\rangle \langle \bar{q}_{k}q_{k}\rangle,
\end{equation}%
with ($i,j,k$) being any permutation of ($u,d,s$).
$C_{i}=\left\langle \bar{q}_{i}q_{i}\right\rangle $ is the quark condensate
of the flavor $i$.
The energy density of quark matter is given by
\begin{eqnarray}
\label{eq:eNJL}
\varepsilon_{\rm{NJL}} &=&\sum\limits_{i = u,d,s}
 {\left[ { - \frac{3}{{{\pi ^2}}}\int_{k_F^i}^\Lambda
  {\sqrt {{k^2} + m_i^{ * 2}} } \;{k^2}dk} \right]}
   + 2{G_S}\left( {C_u^2 + C_d^2 + C_s^2} \right) - 4K{C_u}{C_d}{C_s}
   \nonumber \\
 & &
 + 2{G_V}\left( {n_u^2 + n_d^2 + n_s^2} \right)   - {\varepsilon _0},
\end{eqnarray}%
where $\varepsilon_{0}$ is introduced to set $\varepsilon_{\rm{NJL}}=0$ in
the physical vacuum. In Refs.~\cite{Logo13,Bona12}, an effective bag constant
$B^*$ was introduced since there remains uncertainty in the low-density
normalization of pressure in the NJL model.
The authors of Ref.~\cite{Bona12} varied the free parameter $B^*$ in the range
of $-40$ MeV/fm$^3$ to 50 MeV/fm$^3$, and they found that the hadron-quark
transition density would increase with increasing $B^*$.
In the present work, our choice of $\varepsilon_{0}$ corresponds to
a vanishing pressure in the vacuum.

For the quark matter consisting of quarks ($u$, $d$, and $s$) and
leptons ($e$ and $\mu $) in $\beta $ equilibrium,
the chemical potentials satisfy the relations
$\mu_{s}=\mu_{d}=\mu_{u}+\mu_{e}$ and $\mu_{\mu}=\mu_{e}$.
At zero temperature, the chemical potential of the quark flavor $i$
is defined as $\mu_i =\sqrt{{k_{F}^{i}}^{2}+{m_{i}^{\ast}}^{2}}
+ 4 G_V \, n_i$. The total energy density and pressure in the quark matter
are given by
\begin{eqnarray}
\label{eq:e2}
\varepsilon_{\rm{QP}} &=& \varepsilon_{\rm{NJL}}
  +\sum_{l=e,\mu }\frac{1}{\pi^{2}}
\int_{0}^{k_{F}^{l}}\sqrt{k^{2}+m_{l}^{2}}\ k^{2}dk,
\\
P_{\rm{QP}} &=&\sum_{i=u,d,s,e,\mu }n_{i}\mu_{i}-\varepsilon_{\rm{QP}}.
\label{eq:p2}
\end{eqnarray}

\section{Hadron-quark mixed phase with finite-size effects}
\label{sec:4}

To describe the hadron-quark mixed phase, we employ the Wigner--Seitz
approximation, in which the system is divided into equivalent
and charge-neutral cells. We assume that the coexisting hadronic and
quark phases inside the cell are separated by a sharp interface
and the leptons (electrons and muons) are uniformly distributed
throughout the cell. It has been discussed that the geometrical structure
of the mixed phase may change from droplet to rod, slab, tube, and bubble
with increasing density~\cite{Glen01,Endo06}.
For simplicity, we consider only droplet and bubble
phases in the present study.

Generally, the surface and Coulomb contributions are neglected in the bulk
approximation, where the mixed phase is governed by the
Gibbs conditions. When the finite-size effects are taken into account,
the equilibrium conditions for coexisting hadronic and quark phases
should be derived by minimization of the total energy including the
surface and Coulomb contributions, which are different from the Gibbs
conditions without finite-size effects.
The total energy density of the hadron-quark mixed phase is written as
\begin{eqnarray}
\label{eq:fws}
\varepsilon_{\rm{MP}} &=& u \varepsilon_{\rm{QP}}
  + (1 - u)\varepsilon_{\rm{HP}}
  + \varepsilon_{\rm{surf}} + \varepsilon_{\rm{Coul}} ,
\end{eqnarray}%
where $u=V_{\rm{QP}}/(V_{\rm{QP}}+V_{\rm{HP}})$ is the volume fraction
of the quark phase. The energy densities, $\varepsilon_{\rm{HP}}$
and $\varepsilon_{\rm{QP}}$, are given by Eqs.~(\ref{eq:e1})
and (\ref{eq:e2}), respectively.
The surface and Coulomb energy densities for a spherical cell
are given by
\begin{eqnarray}
{\varepsilon}_{\rm{surf}}
&=& \frac{3 \sigma u_{\rm{in}}}{r},
\label{eq:esurf} \\
{\varepsilon}_{\rm{Coul}}
&=& \frac{e^2}{5}
    \left(\delta n_c\right)^{2}r^{2}
    u_{\rm{in}} D\left( u_{\rm{in}}\right),
\label{eq:ecoul}
\end{eqnarray}%
where%
\begin{eqnarray}
\label{eq:Du}
D\left( u_{\rm{in}}\right)
=1-\frac{3}{2}u_{\rm{in}}^{1/3}+\frac{1}{2}u_{\rm{in}}.
\end{eqnarray}%
Here, $u_{\rm{in}}$ denotes the volume fraction of the inner part
with radius $r$, i.e., $u_{\rm{in}}=u$ for droplets and
$u_{\rm{in}}=1-u$ for bubbles.
$\sigma$ is the surface tension of the hadron-quark interface,
which is treated as a free parameter in the present calculation.
$\delta n_c=n_c^{\rm{HP}}-n_c^{\rm{QP}}$ is the charge-density
difference between the hadronic and quark phases.
The energy density of the mixed phase $\varepsilon_{\rm{MP}}$ can be
considered as a function of nine variables:
$n_{p}$, $n_{n}$, $n_{u}$, $n_{d}$, $n_{s}$, $n_{e}$, $n_{\mu}$, $u$, and $r$.
We derive the equilibrium conditions by minimizing $\varepsilon_{\rm{MP}}$
under the constraints of global charge neutrality and fixed average baryon
density $n_b$, which are written as
\begin{eqnarray}
\label{eq:nc}
0 &=& \frac{u}{3}\left( 2n_u - n_d - n_s \right)
  + (1 - u)  n_p  - n_e - n_\mu , \\
n_b &=& \frac{u}{3}\left( n_u + n_d + n_s \right)
    + (1 - u) \left( n_p + n_n \right).
\label{eq:nb}
\end{eqnarray}
By introducing the Lagrange multipliers, $\mu_e$ and $\mu_n$,
for these two constraints, we perform the minimization for the function
\begin{eqnarray}
w &=&\varepsilon_{\rm{MP}}
  -\mu_n \left[ \frac{u}{3}\left( n_u + n_d + n_s \right)
    + (1 - u) \left( n_p + n_n \right)\right] \notag \\
  & & -\mu_e \left[ n_e + n_\mu - \frac{u}{3}\left( 2n_u - n_d - n_s \right)
    - (1 - u)  n_p  \right].
\end{eqnarray}
Minimizing $w$ with respect to the particle densities yields
the following equilibrium conditions for the chemical potentials:
\begin{eqnarray}
\mu_u - \frac{ 4\varepsilon_{\rm{Coul}} }{3u \, \delta n_c}
  &=& \frac{1}{3}\mu_n - \frac{2}{3}\mu_e, \label{eq:CU}\\
\mu_d + \frac{ 2\varepsilon_{\rm{Coul}} }{3u \, \delta n_c}
  &=& \frac{1}{3}\mu_n + \frac{1}{3}\mu_e, \label{eq:CD}\\
\mu_s + \frac{ 2\varepsilon_{\rm{Coul}} }{3u \, \delta n_c}
  &=& \frac{1}{3}\mu_n + \frac{1}{3}\mu_e, \label{eq:CS}\\
\mu_p +\frac{ 2\varepsilon_{\rm{Coul}} }{(1-u) \, \delta n_c}
  &=& \mu_n - \mu_e, \label{eq:CN}\\
\mu_\mu &=& \mu_e. \label{eq:CE}
\end{eqnarray}
The minimization over $u$ leads to the equilibrium condition for the
pressure
\begin{eqnarray}
P_{\rm{HP}} &=& P_{\rm{QP}} -\frac{2\varepsilon_{\rm{Coul}}}{\delta n_c}
  \left[ \frac{1}{3u}\left( 2n_u - n_d - n_s \right)+\frac{1}{1-u}n_p\right]
  \mp \frac{\varepsilon_{\rm{Coul}} }{u_{\rm{in}}}
  \left(3+u_{\rm{in}}\frac{D^{^{\prime }}}{D}\right),
\label{eq:CP}
\end{eqnarray}
where the sign of the last term is $-$
for droplets and $+$ for bubbles.
The minimization over $r$ results in the equilibrium condition between
surface and Coulomb energies,
\begin{eqnarray}
\varepsilon_{\rm{surf}} &=& 2\varepsilon_{\rm{Coul}},
\label{eq:Csurf}
\end{eqnarray}
which implies that the radius of the droplet or bubble is given by
\begin{eqnarray}
\label{eq:r}
r &=& \left[\frac{15\sigma}
{2 e^2 \left(\delta n_c\right)^{2} D\left( u_{\rm{in}}\right)} \right]^{1/3}.
\end{eqnarray}
It is clear that the equilibrium equations~(\ref{eq:CU})--(\ref{eq:CP})
are different from the Gibbs equilibrium conditions due to the inclusion
of surface and Coulomb energies in the minimization procedure.
However, these equations would reduce to the Gibbs conditions
when the surface and Coulomb energies are neglected.

By solving the above equilibrium equations at a given baryon
density $n_b$, we can obtain the properties of coexisting hadronic
and quark phases, and then calculate thermodynamic quantities of
the mixed phase. The pressure of the mixed phase is extracted from
the thermodynamic relation,
$P_{\rm{MP}} = n_b^{2}\frac{\partial \left(\varepsilon_{\rm{MP}}/n_b\right) }
{\partial n_b}$. Due to the inclusion of surface and Coulomb energies,
$P_{\rm{MP}}$ is no longer equal to $P_{\rm{HP}}$ and $P_{\rm{QP}}$,
which is similar to the case of nuclear liquid-gas phase transition
at subnuclear densities~\cite{Latt91,Bao14b,Baym71}.

\section{Results and discussion}
\label{sec:5}

\begin{figure}[htb]
\includegraphics[bb=40 5 580 400, width=7 cm,clip]{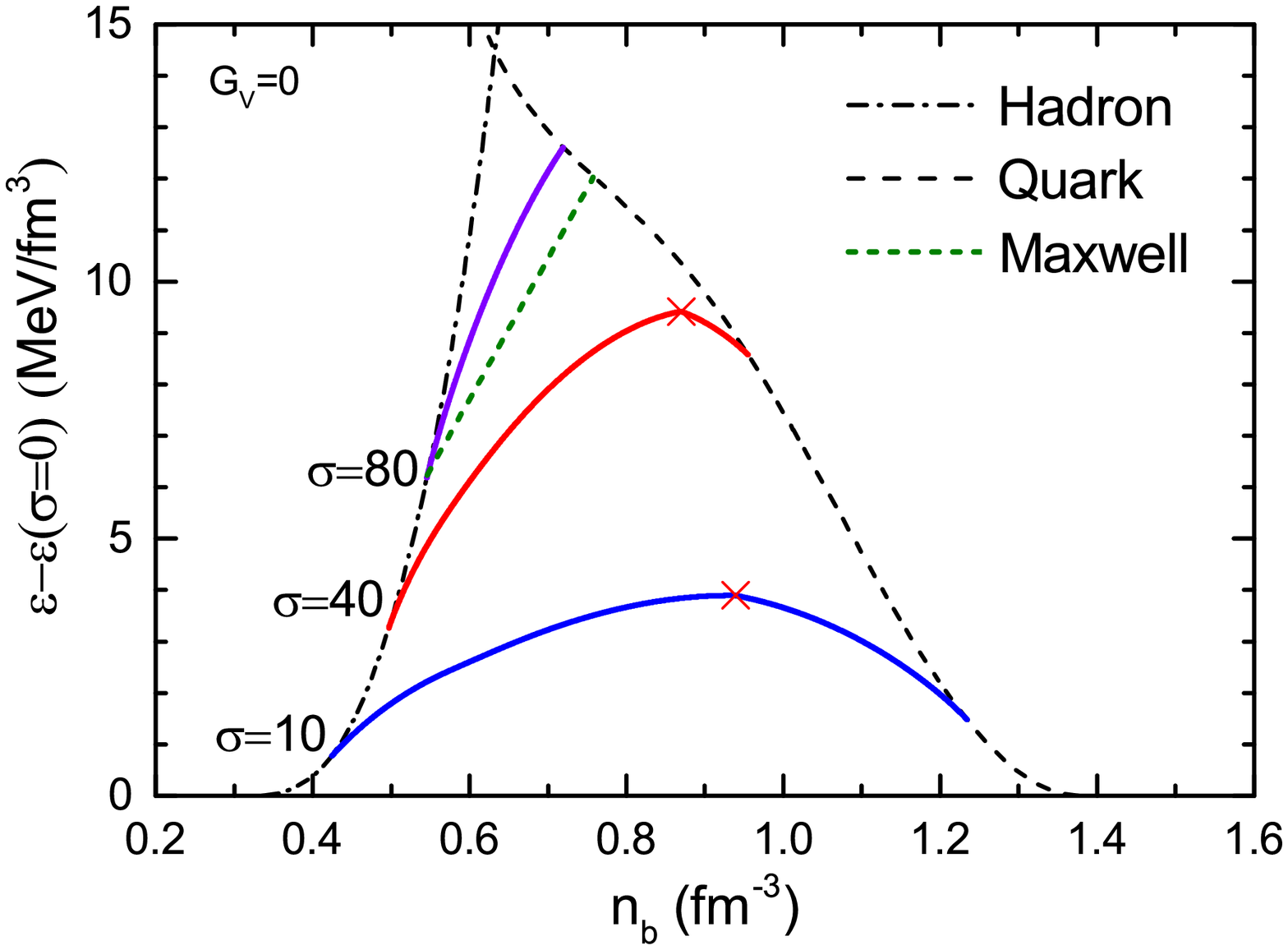}
\includegraphics[bb=40 5 580 400, width=7 cm,clip]{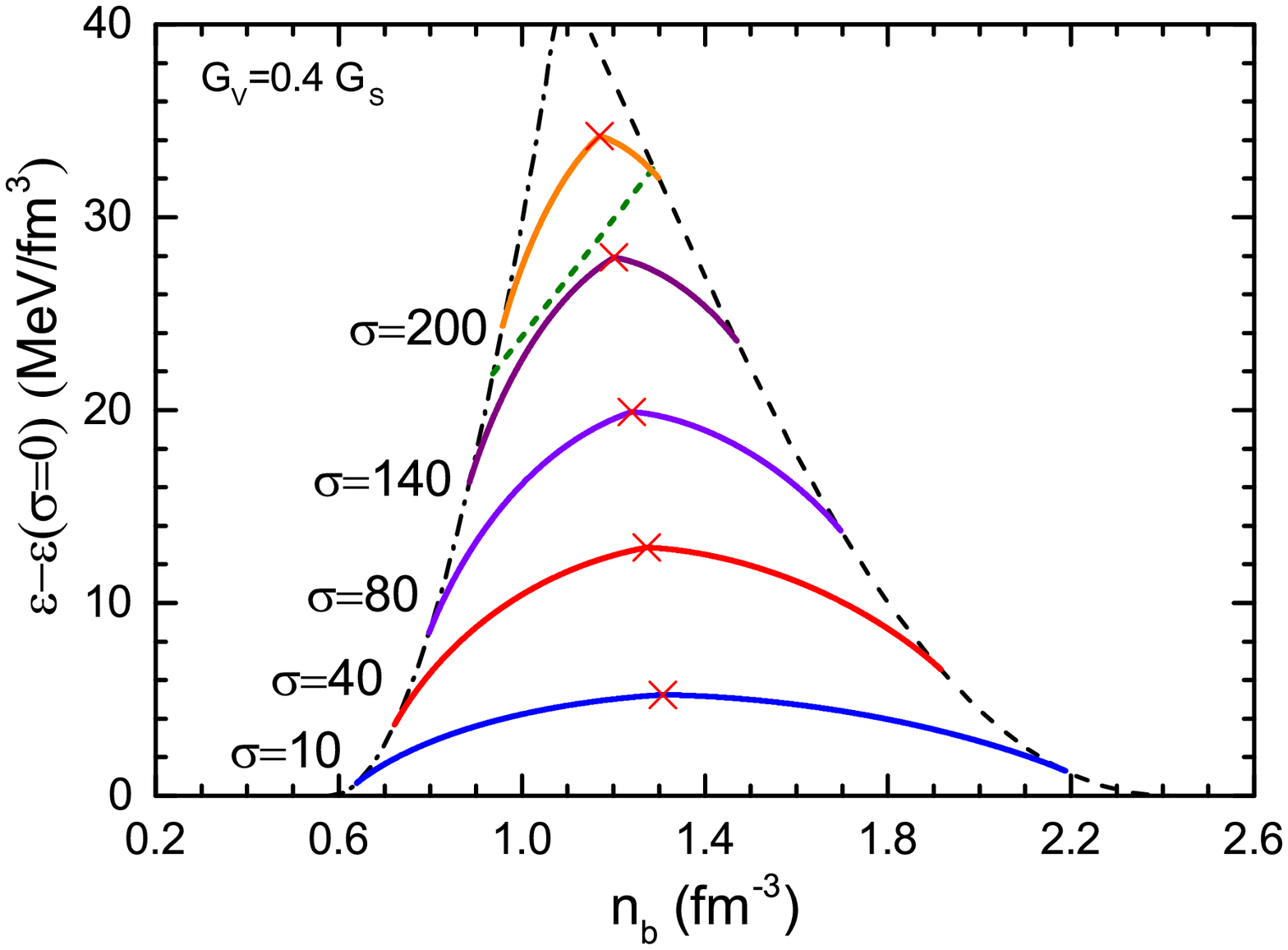}
\caption{(Color online) Energy densities of the mixed phase for
different values of the surface tension $\sigma$,
relative to that of the Gibbs construction
without surface and Coulomb energies ($\sigma=0$).
The cross symbols mark the transition from the droplet phase to the bubble phase.
The results of the Maxwell construction are indicated by the green dotted lines.
The left and right panels correspond to results for $G_V=0$ and $G_V=0.4\,G_S$,
respectively.
}
\label{fig:1enb}
\end{figure}

In this section, we present numerical results for the hadron-quark
phase transition with finite-size effects.
The hadron-quark mixed phase is obtained by solving the equilibrium
conditions under the constraints of global charge neutrality
and baryon number conservation.
We consider both quark droplet and bubble phases in the Wigner-Seitz
approximation. It has been pointed out in Ref.~\cite{Heis93}
that the droplet phase may become energetically unfavorable for large
surface tension ($\sigma > 70$ MeV/fm$^2$), since the energy density
of the droplet phase is higher than those of pure hadronic matter,
pure quark matter, and the mixed phase in the Maxwell construction.
In the present work, we first examine how large the surface tension is
allowed to ensure that the droplet or bubble phase is energetically
favorable. In Fig.~\ref{fig:1enb}, we plot the energy densities of
the mixed phase for various values of the surface tension $\sigma$,
relative to that of the Gibbs construction ($\sigma=0$). The cross
symbols mark the transition from the droplet phase to the bubble phase.
The energy densities of pure hadronic matter and pure quark matter
are shown for comparison. The mixed phase in the Maxwell construction,
which contains locally charge-neutral hadronic and quark matter,
has higher energy density than that of the Gibbs construction,
and their differences are indicated by the green dotted lines.
The results with the vector coupling $G_V=0$ and $G_V=0.4\,G_S$ are
displayed in the left and right panels, respectively.
It is shown that the droplet or bubble phase with $\sigma > 80$ MeV/fm$^2$
($\sigma > 200$ MeV/fm$^2$) for $G_V=0$ ($G_V=0.4\,G_S$) is energetically
unfavorable due to its larger energy density than that of the Maxwell
construction. This implies that the Maxwell construction is preferred
and the local charge neutrality is required for such high surface tension.
In this study, we focus on the difference from the Gibbs construction
caused by surface and Coulomb energies, so we will perform
the calculation for relatively small values of the surface tension.
By comparing the left and right panels of Fig.~\ref{fig:1enb}, we can see that
the density range of the mixed phase for $G_V=0.4\,G_S$ is shifted
to larger value and much wider than that for $G_V=0$. This is because the repulsive
vector interactions in the NJL model can significantly stiffen the EOS of quark matter,
which results in a delay of the phase transition. At higher density, the surface
tension has less impact on the mixed phase, and therefore the allowed values of
the surface tension $\sigma$ for $G_V=0.4\,G_S$ are much larger than that for $G_V=0$.

\begin{figure}[htb]
\includegraphics[bb=40 5 580 580, width=7 cm,clip]{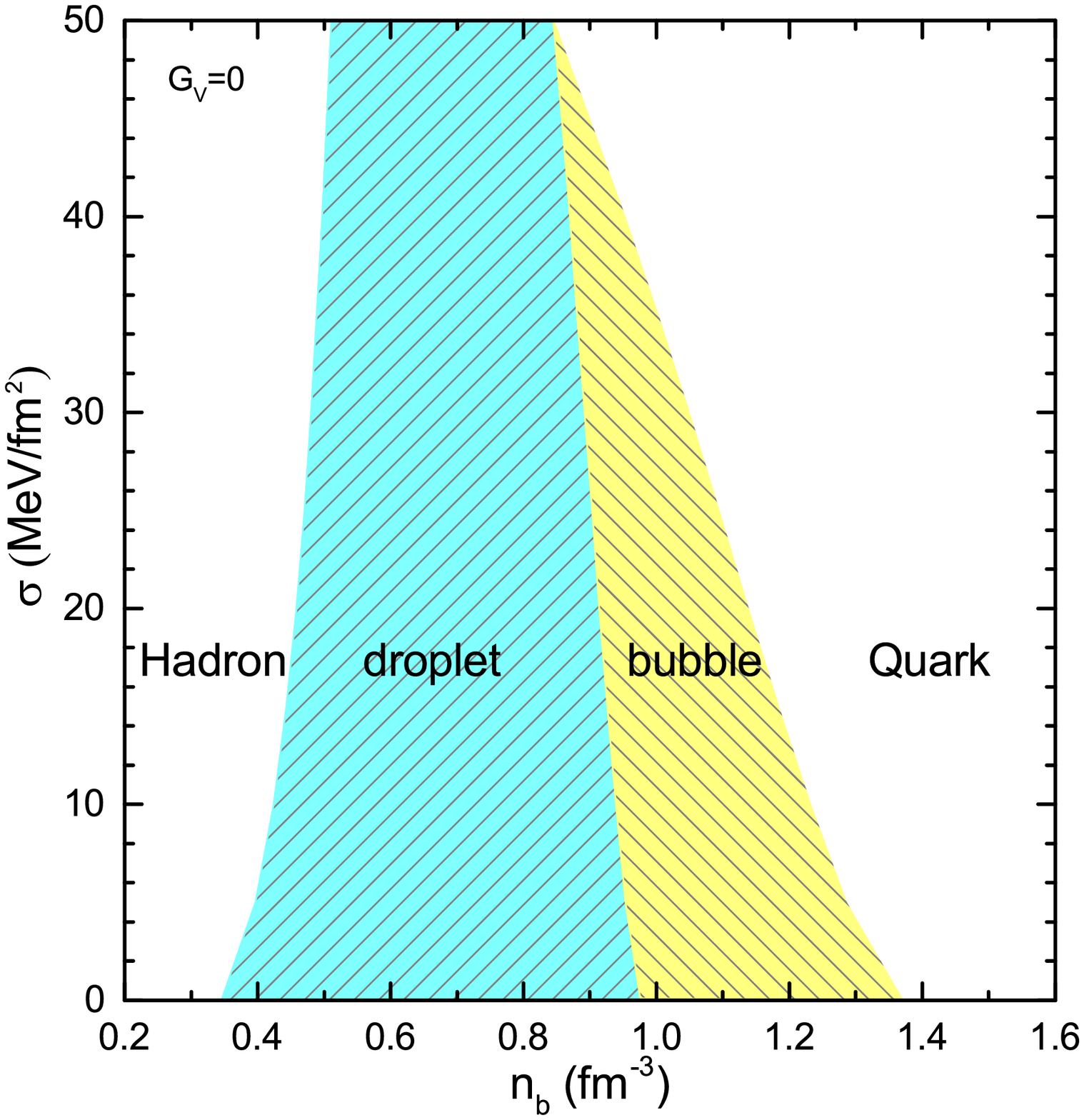}
\includegraphics[bb=40 5 580 580, width=7 cm,clip]{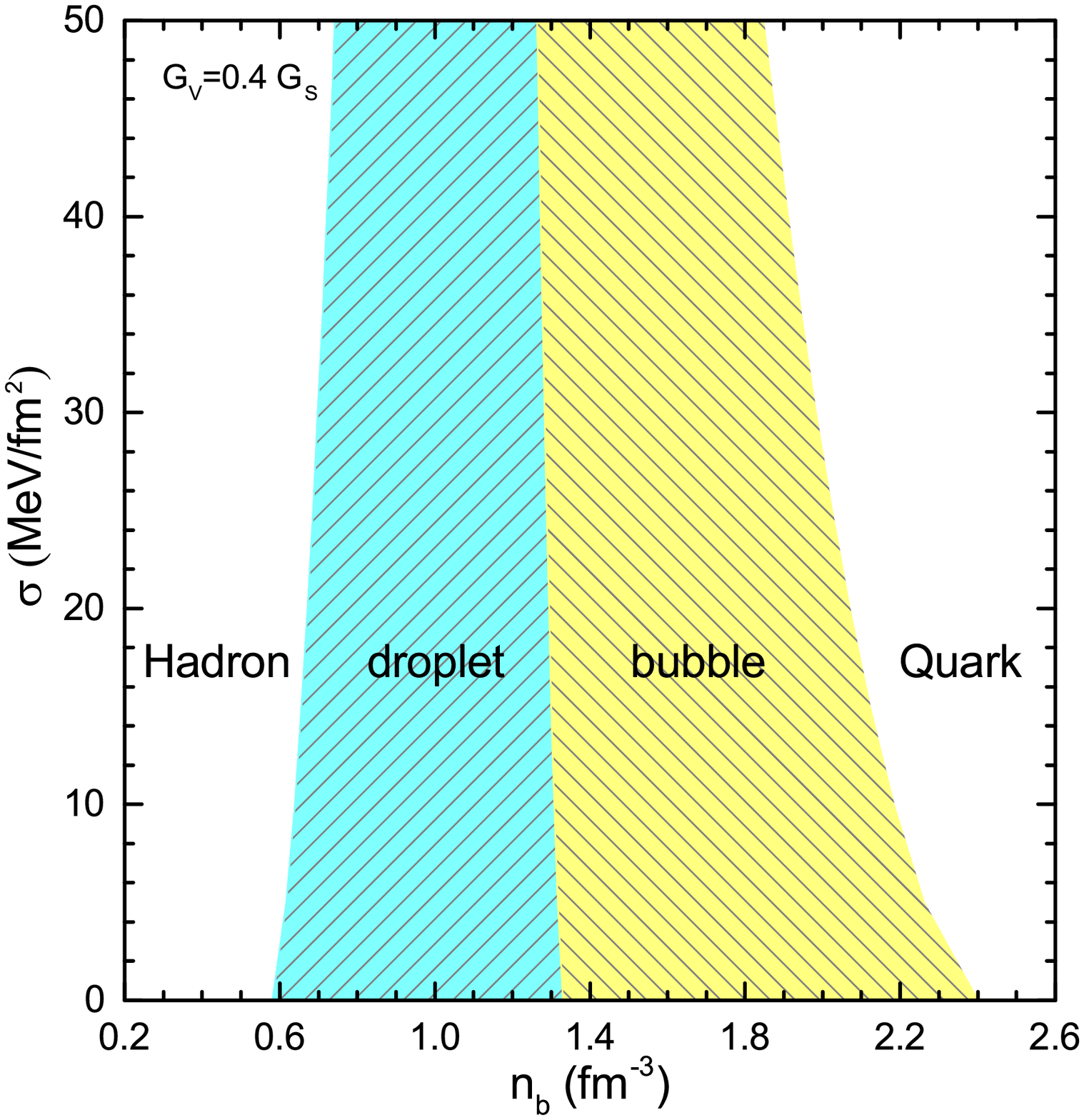}
\caption{(Color online) Phase transition densities as a function
of the surface tension $\sigma$. The shaded region indicates the density
range of the mixed phase in the droplet and bubble configurations.
The left and right panels correspond to results for $G_V=0$ and $G_V=0.4\,G_S$,
respectively.}
\label{fig:2snb}
\end{figure}

In Fig.~\ref{fig:2snb}, we show the density range of the mixed phase
as a function of the surface tension $\sigma$ for $G_V=0$ (left panel)
and $G_V=0.4\,G_S$ (right panel).
At the beginning of the mixed phase, quark matter occupies a small
volume fraction and the favored structure is quark droplets embedded
in hadronic matter. However, toward the end of the mixed phase,
the quark bubble phase is more stable than the droplet phase.
It is known that other geometrical structures, such as rod, slab, and
tube, may exist in the middle of the mixed phase,
which have been neglected in this calculation for simplicity.
As one can see from Fig.~\ref{fig:2snb}, the density range of the
mixed phase is significantly reduced as $\sigma$ increases.
Particularly, the range of the bubble phase gets smaller
and eventually disappears for $\sigma > 50$ MeV/fm$^2$ in the case of $G_V=0$.
Compared to the left panel for $G_V=0$, the density range of the mixed phase
for $G_V=0.4\,G_S$ (shown in the right panel) is shifted to higher densities
and its dependence on $\sigma$ is relatively weak. This is because, as density
increases, the contribution from the surface term becomes less important
relative to the bulk energy. As a result, the influence of the surface tension
$\sigma$ on the phase diagram becomes smaller at higher densities as shown in
the right panel of Fig.~\ref{fig:2snb}.

\begin{figure}[htb]
\includegraphics[bb=35 5 580 580, width=7 cm,clip]{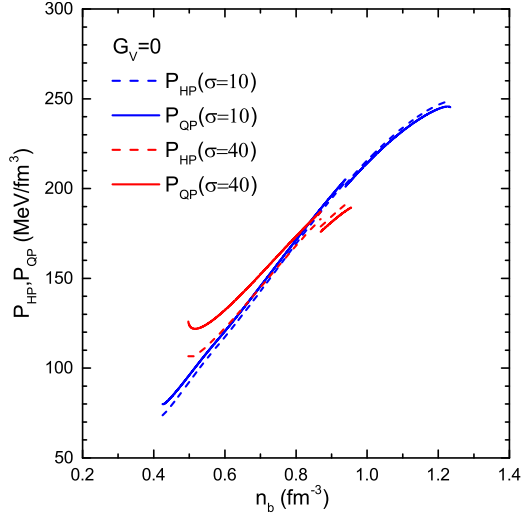}
\caption{(Color online) Pressures of hadronic and quark phases,
$P_{\rm{HP}}$ and $P_{\rm{QP}}$, as a function of the baryon density
in the mixed phase with $\sigma = 10$ and 40 MeV/fm$^2$ for $G_V=0$. }
\label{fig:3pnb}
\end{figure}

\begin{figure}[htb]
\includegraphics[bb=40 5 580 580, width=7 cm,clip]{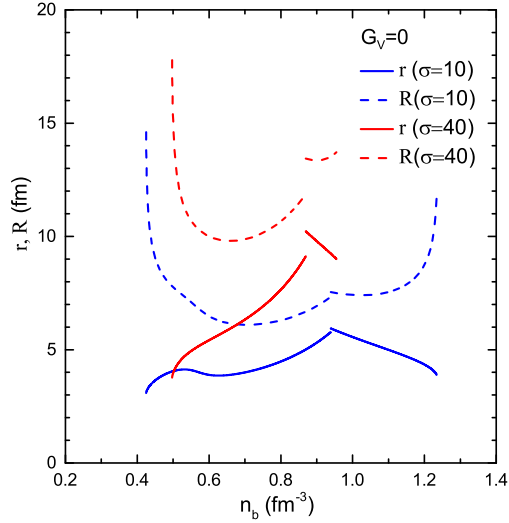}
\caption{(Color online) Radius of the droplet or bubble ($r$)
and that of the Wigner-Seitz cell ($R$) as a function of the baryon
density with $\sigma = 10$ and 40 MeV/fm$^2$ for $G_V=0$.}
\label{fig:4rnb}
\end{figure}

\begin{figure}[htb]
\includegraphics[bb=40 5 580 580, width=7 cm,clip]{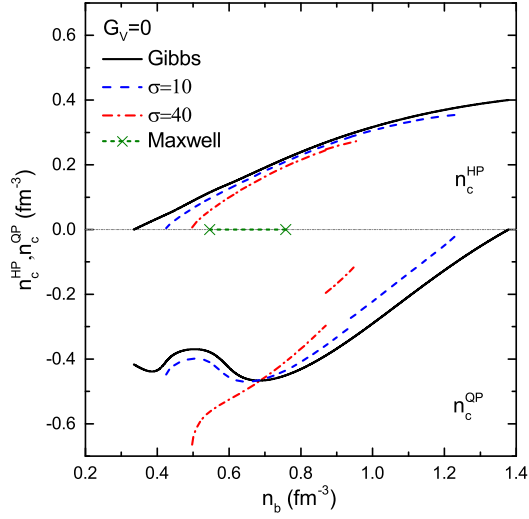}
\caption{(Color online) Charge densities of hadronic
and quark phases, $n_c^{\rm{HP}}$ and $n_c^{\rm{QP}}$, as a function
of the baryon density.
The results with $\sigma=10$ and 40 MeV/fm$^2$ are compared to
those of the Gibbs and Maxwell constructions.}
\label{fig:5ncnb}
\end{figure}

It is interesting to examine the influence of surface and Coulomb
energies on properties of the mixed phase. The Gibbs conditions
for phase equilibrium demand equal pressures and chemical potentials
for coexisting phases. However, when surface and Coulomb energies
are taken into account, the pressure of quark matter is different
from that of hadronic matter, as indicated in Eq.~(\ref{eq:CP}).
In Fig.~\ref{fig:3pnb}, we plot the pressures of hadronic and
quark phases, $P_{\rm{QP}}$ and $P_{\rm{HP}}$, in the mixed phase
obtained with $\sigma = 10$ and 40 MeV/fm$^2$ for $G_V=0$.
It is shown that the differences between $P_{\rm{QP}}$ and $P_{\rm{HP}}$
are very small for $\sigma = 10$ MeV/fm$^2$, while evident differences
are observed for $\sigma = 40$ MeV/fm$^2$, especially at low densities.
The pressures coming from the surface and Coulomb energies have
opposite signs, and the one from the surface tension is somewhat
larger than that from the Coulomb energy. Therefore, the pressure
of the inner phase is slightly higher than that outside.
In Fig.~\ref{fig:4rnb}, we show the radius of the inner part ($r$) and
that of the Wigner-Seitz cell ($R$) as a function of the baryon density $n_b$
obtained with $\sigma = 10$ and 40 MeV/fm$^2$ for $G_V=0$.
As density increases, we can see that $r$ increases in the droplet phase
and then turns to decrease in the bubble
phase, but $R$ shows rather different behavior.
This is related to the increase of the quark volume fraction in the mixed phase.
It is seen that both $r$ and $R$ for $\sigma = 40$ MeV/fm$^2$
are larger than those for $\sigma = 10$ MeV/fm$^2$.
This is because a large value of $\sigma$ favors a large $r$ as
indicated in Eq.~(\ref{eq:r}), and meanwhile, a large $R$ is achieved
according to $R = r u_{\rm{in}}^{-1/3}$.
In Fig.~\ref{fig:5ncnb}, the electric charge densities of hadronic and
quark phases, $n_c^{\rm{HP}}$ and $n_c^{\rm{QP}}$,
are shown as a function of $n_b$ for the same values of
$\sigma$ and $G_V$ as in Figs.~\ref{fig:3pnb} and~\ref{fig:4rnb}.
The Gibbs construction corresponds to $\sigma=0$, which contains positively
charged hadronic matter and negatively charged quark matter with relatively
large differences between $n_c^{\rm{HP}}$ and $n_c^{\rm{QP}}$.
In contrast, the Maxwell construction consists of
two charge-neutral phases, i.e., $n_c^{\rm{HP}}=n_c^{\rm{QP}}=0$,
which is caused by extremely high surface tension.
The results obtained with $\sigma = 10$ and 40 MeV/fm$^2$ are
somewhat different from those of the Gibbs construction,
and a larger value of $\sigma$ results in more significant differences.
In Figs.~\ref{fig:3pnb}--\ref{fig:5ncnb}, we show results only for $G_V=0$;
however, similar behaviors are observed for other values of $G_V$.

\begin{center}
\begin{figure}[htb]
\centering
\begin{tabular}{ccc}
\includegraphics[bb=20 8 570 400, width=0.33\linewidth, clip]{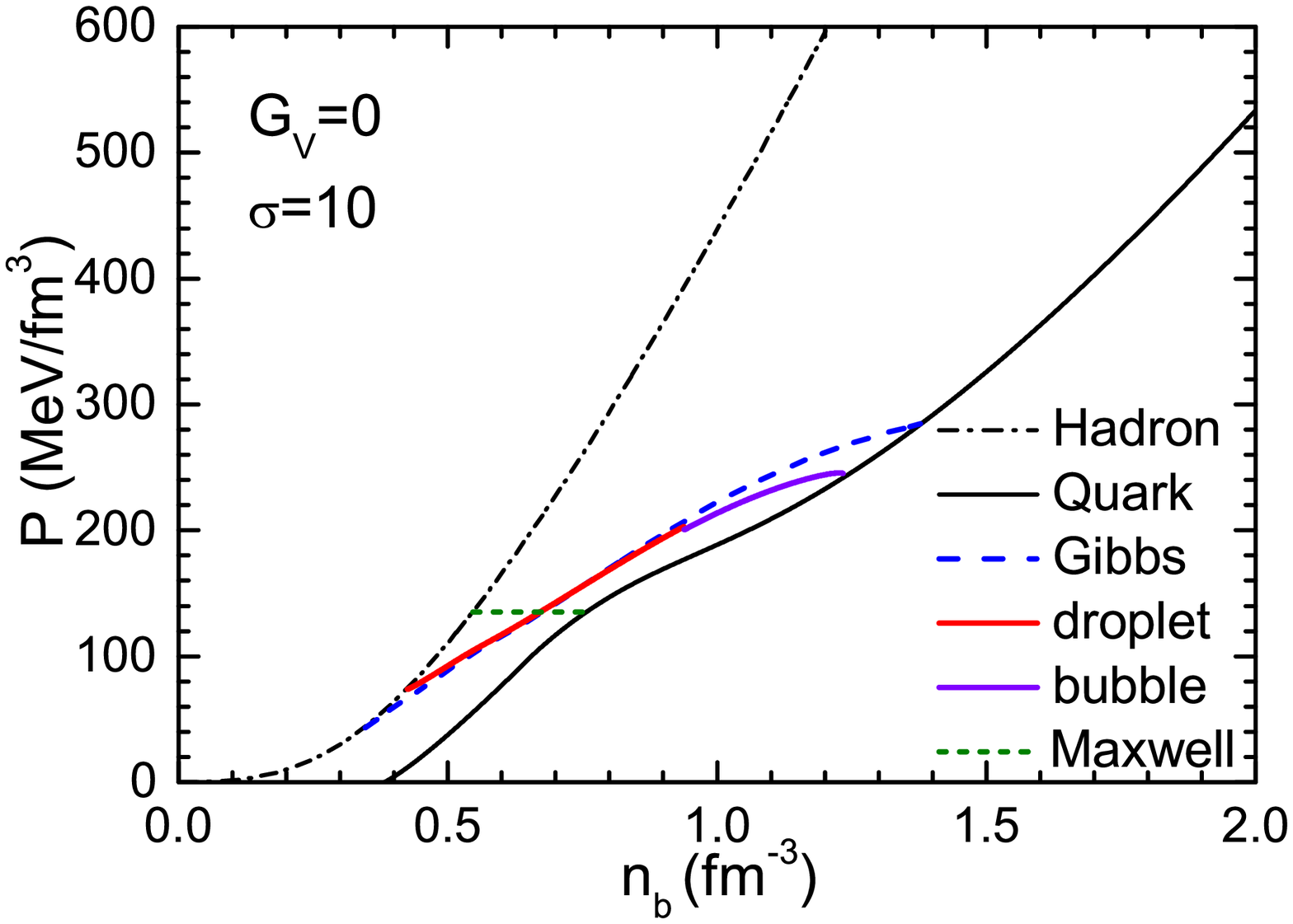}&
\includegraphics[bb=20 8 570 400, width=0.33\linewidth, clip]{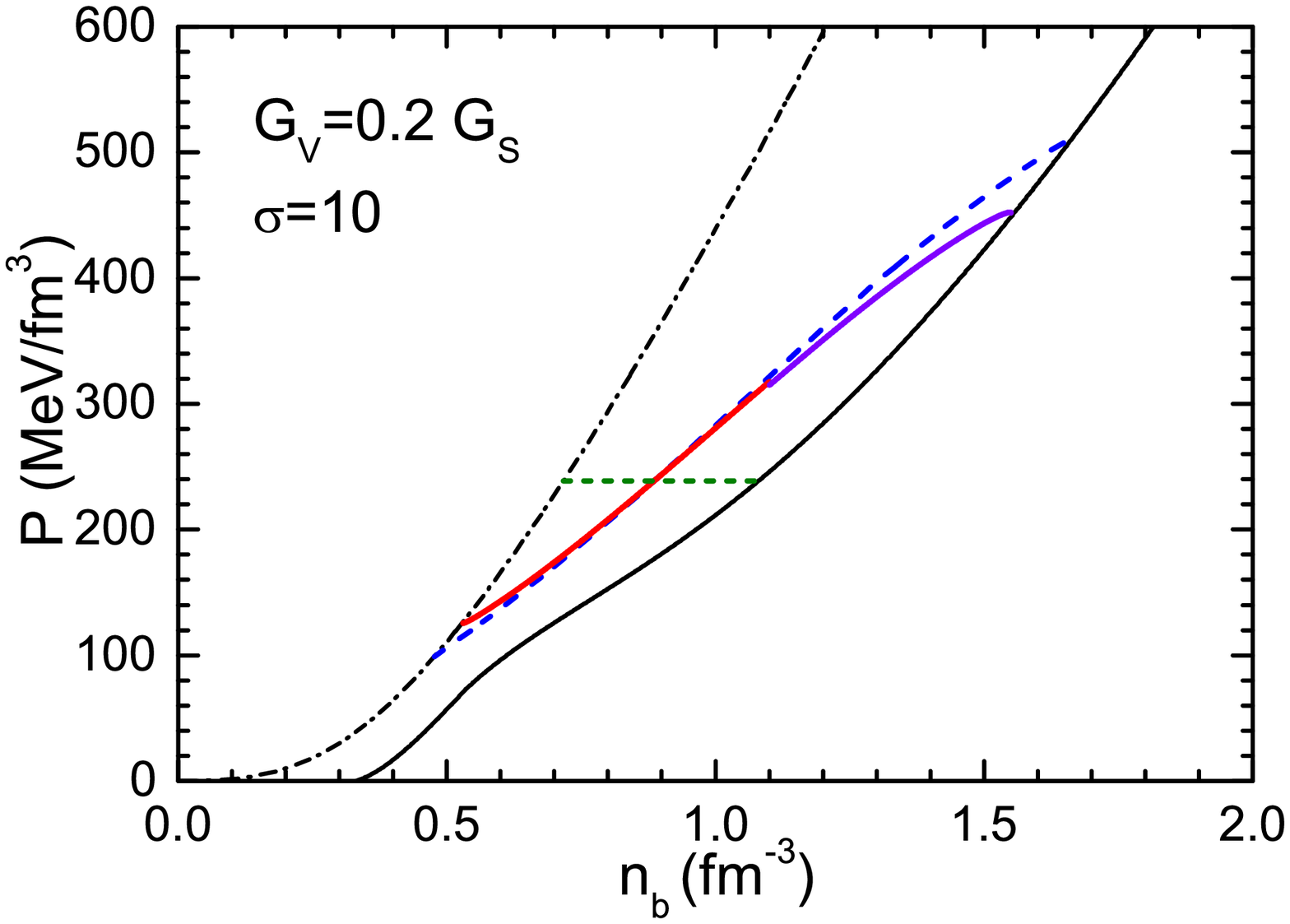}&
\includegraphics[bb=20 8 570 400, width=0.33\linewidth, clip]{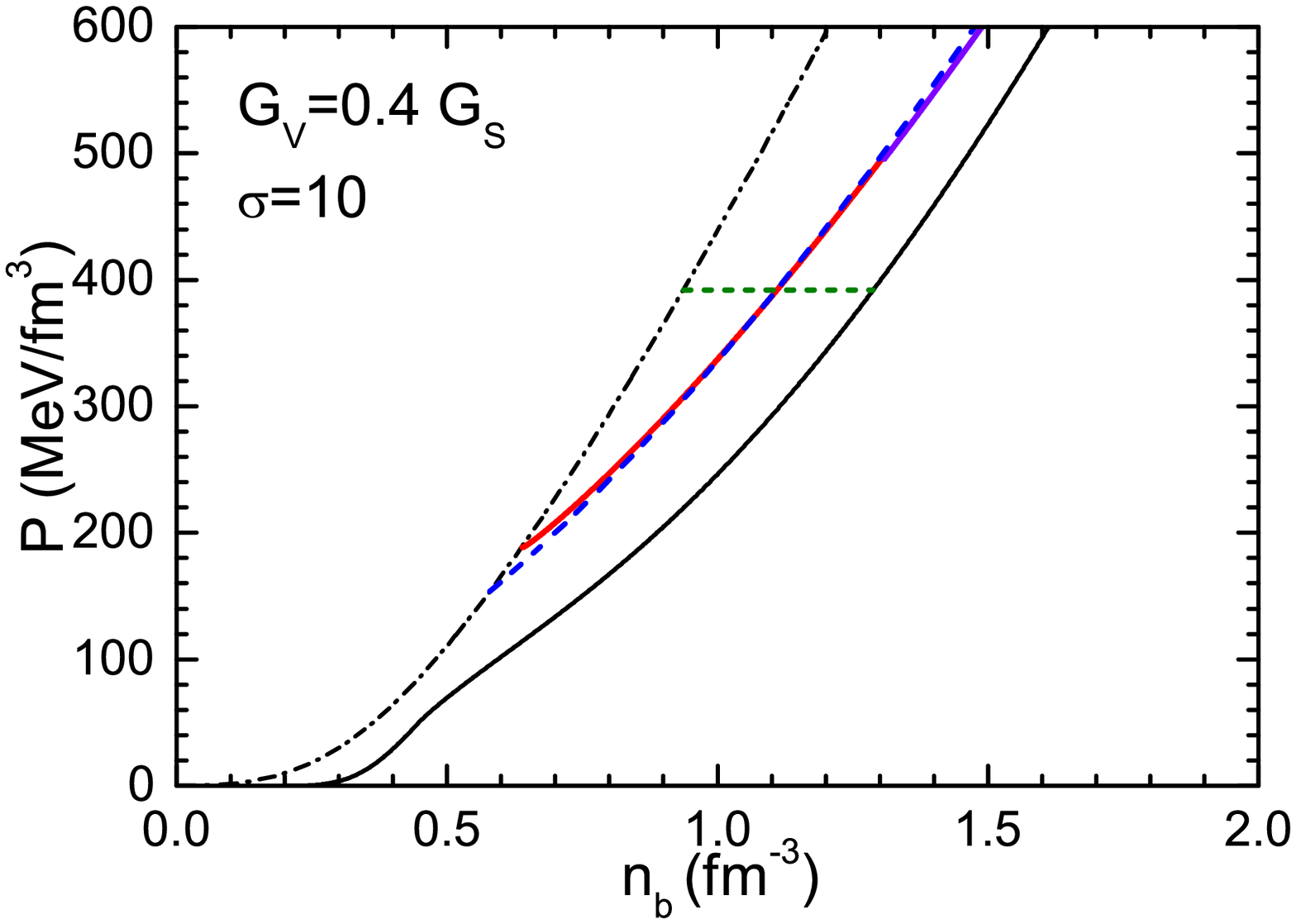} \\
\includegraphics[bb=20 8 570 400, width=0.33\linewidth, clip]{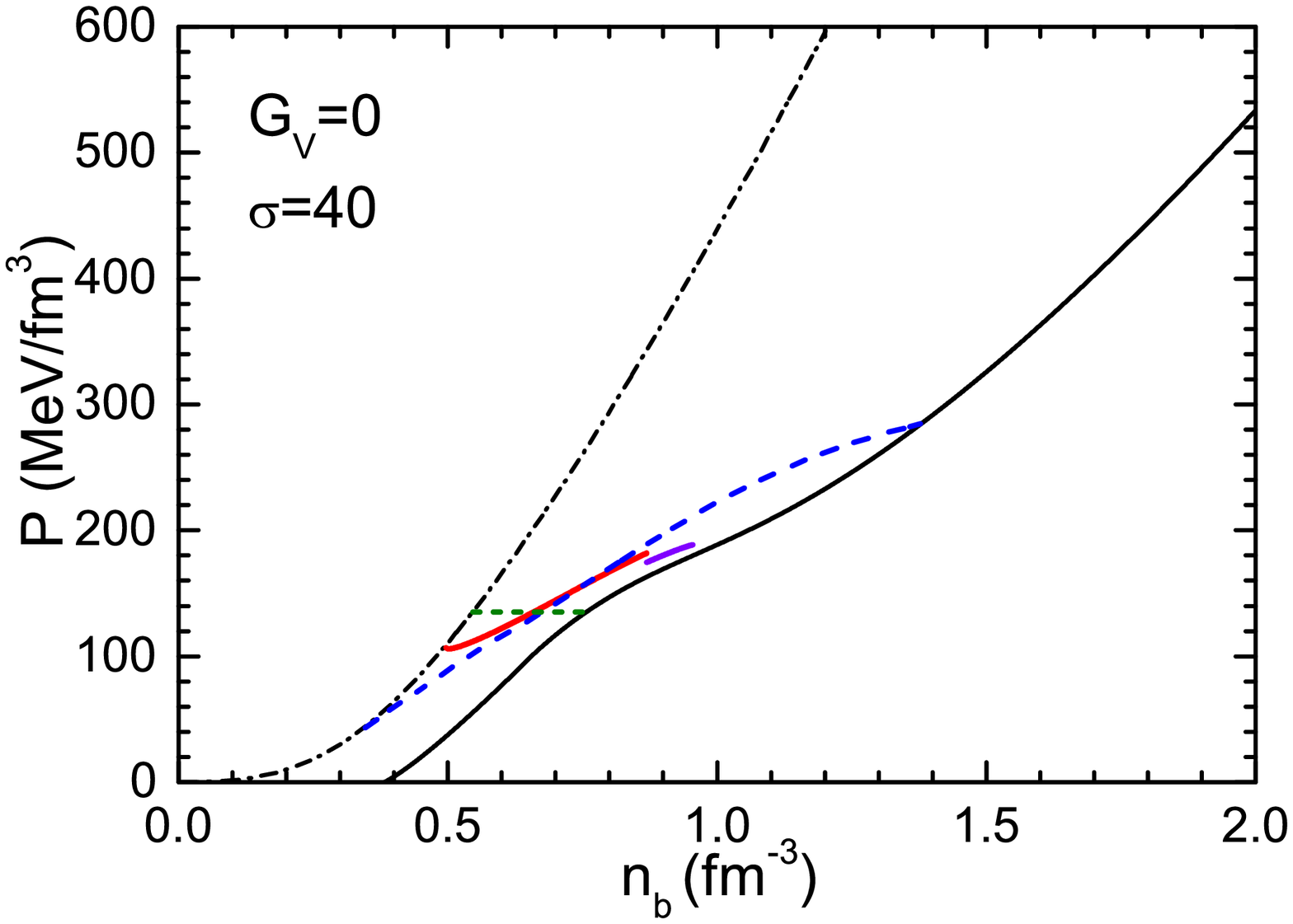}&
\includegraphics[bb=20 8 570 400, width=0.33\linewidth, clip]{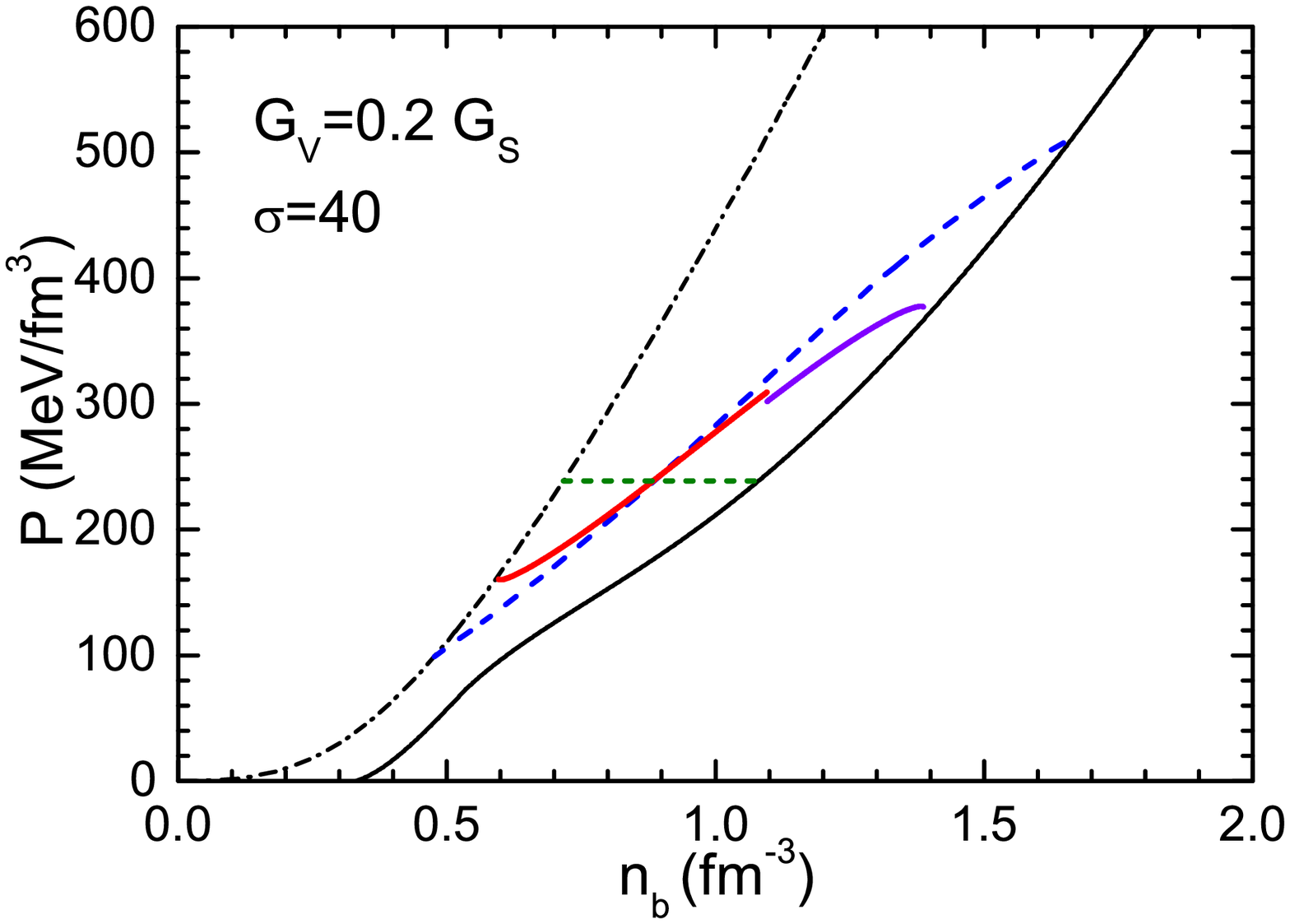}&
\includegraphics[bb=20 8 570 400, width=0.33\linewidth, clip]{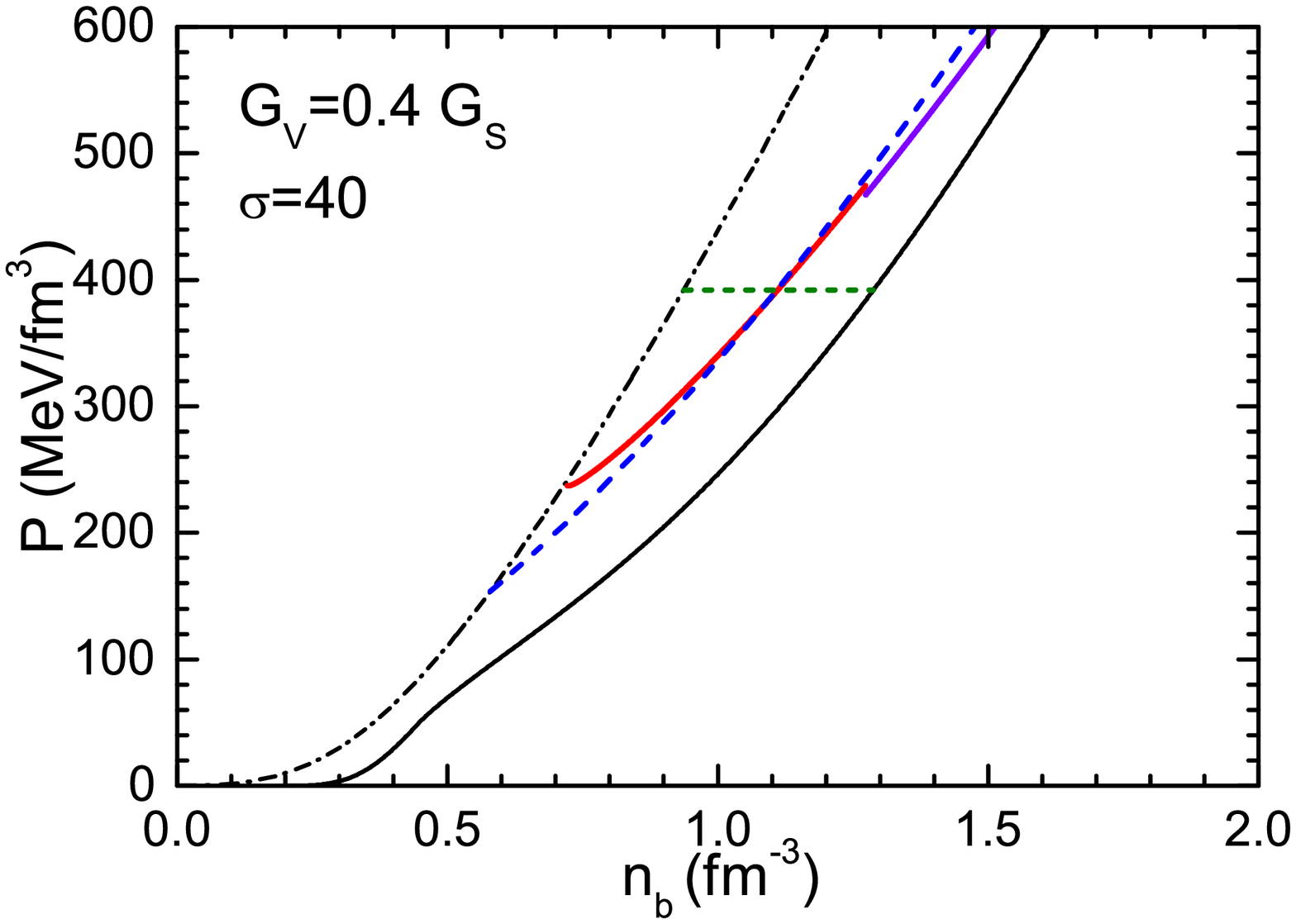} \\
\end{tabular}
\caption{(Color online) Pressures as a function of the baryon density
for hadronic, mixed, and quark phases.
The results of the mixed phase with $\sigma=10$ MeV/fm$^2$ (upper panel)
and $\sigma=40$ MeV/fm$^2$ (lower panel)
are compared to those of the Gibbs and Maxwell constructions. The results
for $G_V=0$, $G_V=0.2\,G_S$, and $G_V=0.4\,G_S$ are shown in the left,
middle, and right panels, respectively.}
\label{fig:6pnb}
\end{figure}
\end{center}

In Fig.~\ref{fig:6pnb}, we plot the pressures as a function of the baryon
density for hadronic, mixed, and quark phases.
The left, middle, and right panels show respectively the results for $G_V=0$,
$0.2\,G_S$, and $0.4\,G_S$, while the upper and lower panels correspond to the results
of the mixed phase obtained with $\sigma = 10$ and 40 MeV/fm$^2$.
The droplet and bubble phases are indicated by the red and purple solid lines.
For comparison, results with the Gibbs and Maxwell constructions
are shown by the blue dashed and green dotted lines, respectively.
It is shown that pressures of the mixed phase obtained with a finite value
of $\sigma$ lie between those of the Gibbs and Maxwell constructions.
The results of $\sigma = 10$ MeV/fm$^2$ (upper panels) are closer to that of the
Gibbs construction than those of $\sigma = 40$ MeV/fm$^2$ (lower panels).
By comparing the left, middle, and right panels, one can see the effect of the
repulsive vector interactions in the NJL model. As the vector coupling $G_V$
increases, the EOS of quark matter gets stiffer. As a result,
the mixed phase exists in a broad density range and moves toward higher densities.

\begin{figure}[htb]
\includegraphics[bb=35 5 580 580, width=7 cm,clip]{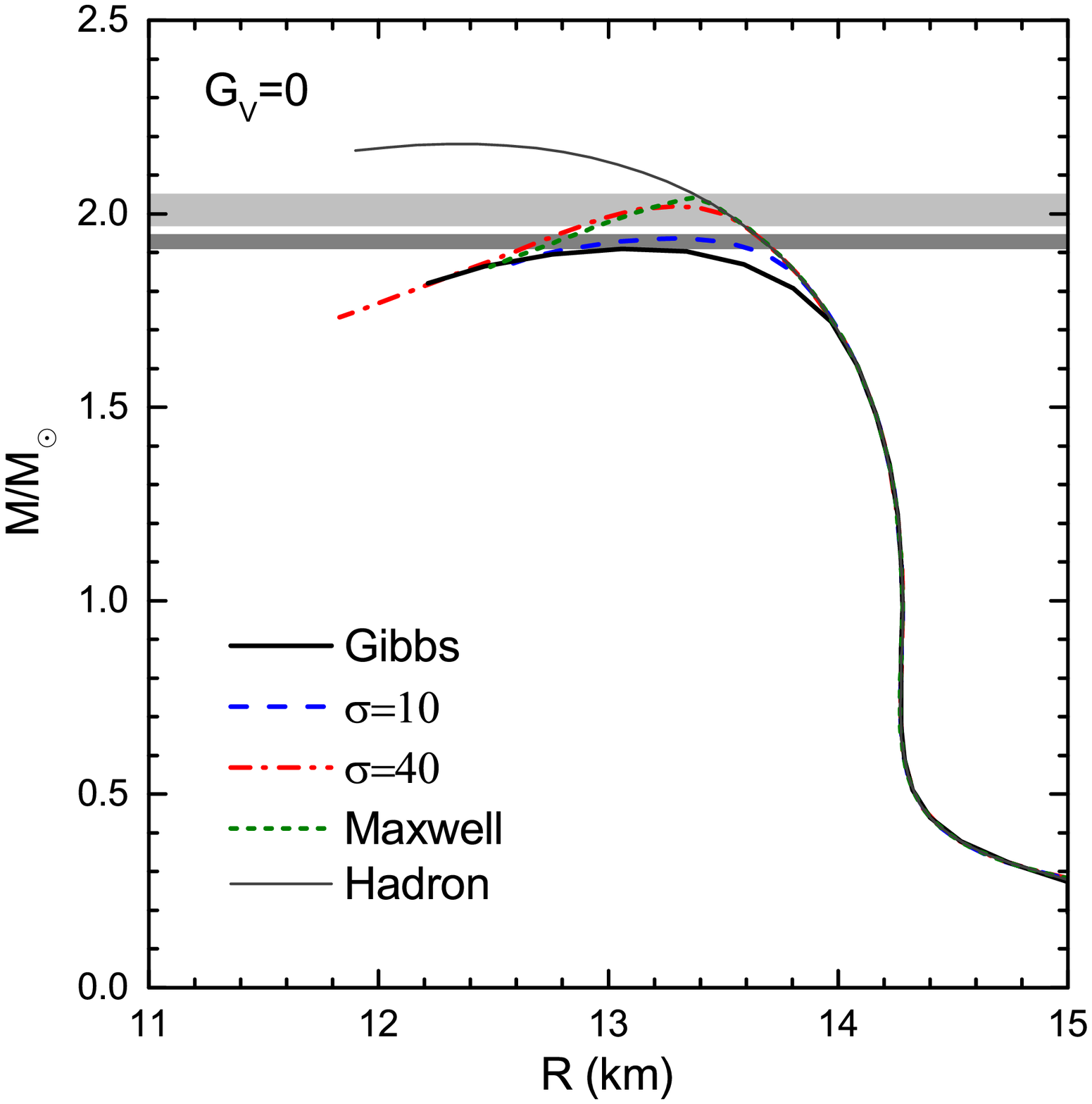}
\includegraphics[bb=35 5 580 580, width=7 cm,clip]{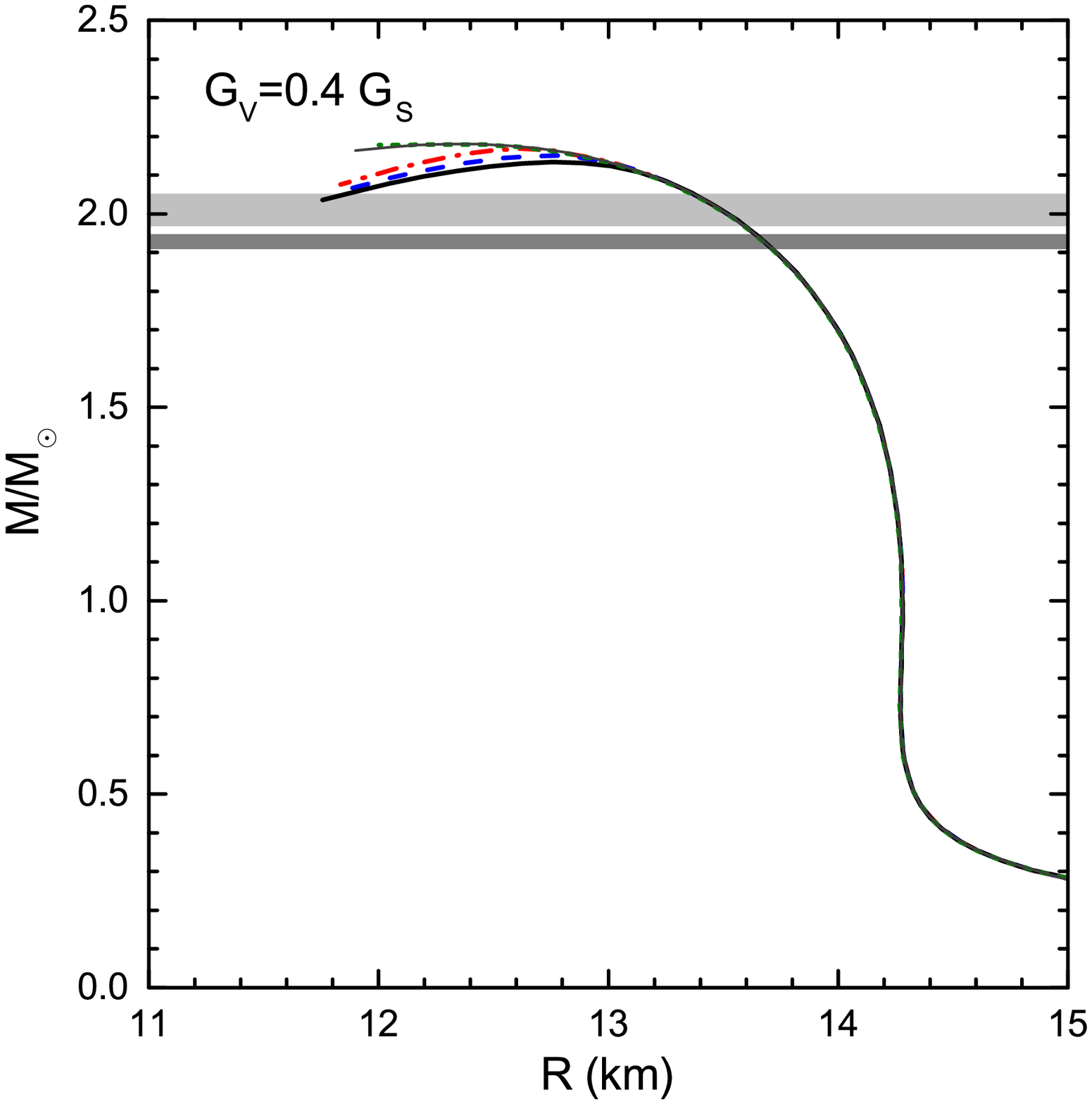}
\caption{(Color online) Mass-radius relations of neutron stars
for different EOS. For comparison, the results from a pure hadronic
EOS are shown by the thin solid lines. The lighter and darker shaded
regions correspond to the observational constraints of PSR J0348--0432
($M=2.01  \pm 0.04  \ M_\odot$)~\cite{Anto13} and PSR J1614--2230
($M=1.928 \pm 0.017 \ M_\odot$)~\cite{Fons16}, respectively.}
\label{fig:7mr}
\end{figure}

To examine the finite-size effects on properties of neutron stars, we solve
the Tolman-Oppenheimer-Volkoff equation by using the EOS described above
for $G_V=0$ (left panel) and $G_V=0.4\,G_S$ (right panel).
For the description of neutron-star crusts, the present EOS is matched
to the EOS at subnuclear densities, which was calculated from the Thomas-Fermi
approximation by using the TM1 model for nuclear interactions~\cite{Shen02}.
The resulting mass-radius relations are presented in Fig.~\ref{fig:7mr},
where the observational constraints of PSR J0348--0432
($M=2.01  \pm 0.04  \ M_\odot$)~\cite{Anto13} and PSR J1614--2230
($M=1.928 \pm 0.017 \ M_\odot$)~\cite{Fons16} are shown by the lighter
and darker shaded regions, respectively.
For comparison, results of pure hadronic EOS are shown by thin
solid lines, which give a maximum mass of $2.18\ M_{\odot}$~\cite{Shen11}.
The inclusion of quark degrees of freedom significantly softens the EOS
and reduces the maximum mass of neutron stars, which depends on the vector
coupling $G_V$, as shown in the two panels of Fig.~\ref{fig:7mr}.
In the case of $G_V=0.4\,G_S$ ($G_V=0$), the maximum mass with the Gibbs construction
is reduced to $2.13\ M_{\odot}$ ($1.91\ M_{\odot}$).
When the finite-size effects are taken into account,
neutron-star masses are somewhat higher than those of the Gibbs construction
and the differences depend on the surface tension $\sigma$.
In Table~\ref{tab:2}, the calculated properties of neutron stars with the maximum
mass are presented in detail. For the cases of $\sigma=0$ (Gibbs), 10, and 40 MeV/fm$^2$,
a mixed-phase core with radius $R_{\mathrm{MP}}$ can be formed in the interior
of stars and $R_{\mathrm{MP}}$ decreases with increasing $G_V$,
but the central density is not high enough to generate pure quark matter.
For the Maxwell construction, the mixed phase is not allowed to appear
in stars because of its constant pressure. However,
a small quark phase core may exist with $R_{\mathrm{QP}}=0.82$ km for $G_V=0$
and $R_{\mathrm{QP}}=0.38$ km for $G_V=0.2\,G_S$. We notice that there is no
quark matter in the interior of neutron stars for larger vector coupling
$G_V=0.4\,G_S$, as shown in the last line of Table~\ref{tab:2}.
It is found that the internal structures of neutron stars
are rather sensitive to the values of the surface tension $\sigma$
and the vector coupling $G_V$.

\section{Conclusions}
\label{sec:6}

We have investigated the finite-size effects on the hadron-quark phase
transition, which may occur in the interior of massive neutron stars.
The RMF model has been used to describe the hadronic matter phase,
while the NJL model with vector interactions has been adopted for
the quark matter phase. We have employed the Wigner-Seitz approximation
to describe the hadron-quark mixed phase, where the coexisting hadronic
and quark phases inside the charge-neutral cell are separated by a sharp
interface. We have derived the equilibrium conditions for coexisting hadronic and
quark phases by minimization of the total energy including the surface
and Coulomb contributions.
It has been found that these equilibrium conditions are different from
the Gibbs conditions used in the bulk calculations
due to the inclusion of surface and Coulomb energies.
As a consequence, the pressure of quark matter is no longer equal to
that of hadronic matter, and the differences are more pronounced for
larger values of the surface tension.

The effects of the surface tension $\sigma$ and the vector
coupling $G_V$ on properties of the hadron-quark mixed phase have been
investigated in the present work. For large values of $\sigma$,
the density range of the mixed phase is significantly reduced with
respect to that of the Gibbs construction.
Furthermore, a larger surface tension generally leads to a larger structure
size and smaller charge-density difference between the two phases.
Since the Gibbs and Maxwell constructions correspond, respectively,
to the two limits of zero and infinite surface tension,
results for finite values of the surface tension were found
to lie between these two limits.
The repulsive vector interactions in the NJL model
could stiffen the EOS of quark matter, and as a result, the mixed phase
would exist in a broad density range and move toward higher densities.

The properties of neutron stars have been calculated with the inclusion
of finite-size effects. The maximum masses of neutron stars were found
to depend on both the surface tension $\sigma$ and the vector
coupling $G_V$, which increase with increasing $\sigma$ and $G_V$.
The maximum masses for finite values of $\sigma$ were found to lie between
results of the Gibbs and Maxwell constructions. A mixed-phase core might be
formed in the interior of massive stars, but no pure quark phase could exist
for relatively small surface tension in the present study.
In the case of the Maxwell construction, a small pure quark core could
appear for smaller values of $G_V$. It has been noticed that our results
of neutron stars could be compatible with the observations of
PSR J1614--2230 and PSR J0348--0432.
Finally, we emphasize that the surface tension of the hadron-quark interface
and the vector interaction between quarks play critical roles in determining
behaviors of the hadron-quark phase transition and neutron star
properties. Therefore, better estimates for these quantities
are needed for further studies.

\section*{Acknowledgment}

This work was supported in part by the National Natural
Science Foundation of China (Grants No. 11375089 and No. 11675083).


\newpage
\begin{table}[tbp]
\caption{Parameter set TM1 for the RMF Lagrangian. The masses are given in MeV.}
\begin{center}
\begin{tabular}{lccccccccccc}
\hline\hline
Model   &$M$  &$m_{\sigma}$  &$m_\omega$  &$m_\rho$  &$g_\sigma$  &$g_\omega$
        &$g_\rho$ &$g_{2}$ (fm$^{-1}$) &$g_{3}$ &$c_{3}$ \\
\hline
TM1     &938.0  &511.198  &783.0  &770.0  &10.0289  &12.6139  &4.6322
        &$-$7.2325   &0.6183   &71.3075  \\
\hline\hline
\end{tabular}
\label{tab:1}
\end{center}
\end{table}

\begin{table}[tbp]
\caption{Properties of neutron stars with the maximum mass $M_{\mathrm{max}}$.
The central energy density and baryon number density are denoted
by $\varepsilon_c$ and $n_{c}$, respectively.
$R_{\mathrm{QP}}$, $R_{\mathrm{MP}}$, and $R$ correspond to radii of
the quark phase, the mixed phase, and the whole star.}
\label{tab:2}
\begin{center}
\begin{tabular}{llcccccccc}
\hline\hline
    &   & $M_\mathrm{max}$ & $\varepsilon_c$ & $n_c$
 & $R_\mathrm{QP}$ & $R_\mathrm{MP}$ & $R$  \\
    &   &$(M_\odot)$ & $(\rm{MeV/fm}^3)$ & $(\rm{fm}^{-3})$
 &(km) & (km) & (km) \\
\hline
Gibbs                    & $G_V=0$        & 1.91 & 876.3  & 0.76 & $-$  & 7.80 & 13.09 \\
                         & $G_V=0.2\,G_S$ & 2.05 & 912.4  & 0.77 & $-$  & 5.60 & 13.00 \\
                         & $G_V=0.4\,G_S$ & 2.13 & 963.9  & 0.80 & $-$  & 4.50 & 12.77 \\
\hline
$\sigma=10$ MeV/fm$^2 $  & $G_V=0$        & 1.94 & 798.2  & 0.70 & $-$  & 5.60 & 13.30 \\
                         & $G_V=0.2\,G_S$ & 2.08 & 907.3  & 0.77 & $-$  & 4.50 & 13.01 \\
                         & $G_V=0.4\,G_S$ & 2.15 & 948.7  & 0.79 & $-$  & 3.41 & 12.77 \\
\hline
$\sigma=40$ MeV/fm$^2 $  & $G_V=0$        & 2.00 & 792.4  & 0.69 & $-$  & 3.64 & 13.37 \\
                         & $G_V=0.2\,G_S$ & 2.11 & 889.0  & 0.75 & $-$  & 2.95 & 13.03 \\
                         & $G_V=0.4\,G_S$ & 2.17 & 981.5  & 0.81 & $-$  & 2.26 & 12.67 \\
\hline
Maxwell                  & $G_V=0$        & 2.04 & 896.1  & 0.77 & 0.82 & $-$  & 13.40 \\
                         & $G_V=0.2\,G_S$ & 2.16 & 1395.3 & 1.08 & 0.38 & $-$  & 12.77 \\
                         & $G_V=0.4\,G_S$ & 2.18 & 1081.2 & 0.87 & $-$  & $-$  & 12.30 \\

\hline\hline
\end{tabular}
\end{center}
\end{table}

\newpage

\newpage

\begin{thebibliography}{99}

\bibitem{Glen01} N. K. Glendenning,
Phys. Rep. \textbf{342}, 393 (2001).

\bibitem{Heis00} H. Heiselberg and M. Hjorth-Jensen,
Phys. Rep. \textbf{328}, 237 (2000).

\bibitem{Webe05} F. Weber,
Prog. Part. Nucl. Phys. \textbf{54}, 193 (2005).

\bibitem{Glen92} N. K. Glendenning, Phys. Rev. D \textbf{46}, 1274 (1992).

\bibitem{Sche99} K. Schertler, S. Leupold, and J. Schaffner-Bielich,
Phys. Rev. C \textbf{60}, 025801 (1999).

\bibitem{Sche00} K. Schertler, C. Greiner, J. Schaffner-Bielich,
and M. H. Thoma, Nucl. Phys. A \textbf{677}, 463 (2000).

\bibitem{Latt00} A. W. Steiner, M. Prakash, and  J. M. Lattimer,
Phys. Lett. B \textbf{486}, 239 (2000).

\bibitem{Burg02}  G. F. Burgio, M. Baldo, P. K. Sahu, and H.-J. Schulze,
Phys. Rev. C \textbf{66}, 025802 (2002).

\bibitem{Mene03}  D. P. Menezes and C. Provid\^{e}ncia,
Phys. Rev. C \textbf{68}, 035804 (2003).

\bibitem{Shar07}  B. K. Sharma, P. K. Panda, and S. K. Patra,
Phys. Rev. C \textbf{75}, 035808 (2007).

\bibitem{Yang08}  F. Yang and H. Shen,
Phys. Rev. C \textbf{77}, 025801 (2008).

\bibitem{Xu10} J. Xu, L. W. Chen, C. M. Ko, and B. A. Li,
Phys. Rev. C \textbf{81}, 055803 (2010).

\bibitem{Chen13} H. Chen, G. F. Burgio, H.-J. Schulze, and N. Yasutake,
Astron. Astrophys. \textbf{551}, A13 (2013).

\bibitem{Orsa14} M. Orsaria, H. Rodrigues, F. Weber, and G. A. Contrera,
Phys. Rev. C \textbf{89}, 015806 (2014).

\bibitem{Bhat10} A. Bhattacharyya, I. N. Mishustin, and W. Greiner,
J. Phys. G \textbf{37}, 025201 (2010).

\bibitem{Heis93} H. Heiselberg, C. J. Pethick, and E. F. Staubo,
Phys. Rev. Lett. \textbf{70}, 1355 (1993).

\bibitem{Endo06} T. Endo, T. Maruyama, S. Chiba, and T. Tatsumi,
Prog. Theor. Phys. \textbf{115}, 337 (2006).

\bibitem{Maru07} T. Maruyama, S. Chiba, H.-J. Schulze, and T. Tatsumi,
Phys. Rev. D \textbf{76}, 123015 (2007).

\bibitem{Yasu14} N. Yasutake, R. {\L}astowiecki, S. Beni{\'{c}}, D. Blaschke, T. Maruyama,
and T. Tatsumi, Phys. Rev. C \textbf{89}, (2014) 065803

\bibitem{Glen95} N. K. Glendenning and S. Pei,
Phys. Rev. C \textbf{52}, 2250 (1995).

\bibitem{Glen97} M. B. Christiansen and N. K. Glendenning,
Phys. Rev. C \textbf{56}, 2858 (1997).

\bibitem{Vosk05} D. N. Voskresensky, M. Yasuhira, and T. Tatsumi,
Nucl. Phys. A \textbf{723}, 291 (2005).

\bibitem{Latt91} J. M. Lattimer and F. D. Swesty,
Nucl. Phys. A \textbf{535}, 331 (1991).

\bibitem{Bao14b} S. S. Bao, J. N. Hu, Z. W. Zhang, and H. Shen,
Phys. Rev. C \textbf{90}, 045802 (2014).

\bibitem{Berg87} M. S. Berger and R. L. Jaffe,
Phys. Rev. C \textbf{35}, 213 (1987); \textbf{44}, 566(E) (1991).

\bibitem{Lugo13} G. Lugones, A. G. Grunfeld, and M. A. Ajmi,
Phys. Rev. C \textbf{88}, 045803 (2013).

\bibitem{Pint12} M. B. Pinto, V. Koch, and J. Randrup,
Phys. Rev. C \textbf{86}, 025203 (2012).

\bibitem{Hats94} T. Hatsuda and T. Kunihiro, Phys. Rep. \textbf{247}, 221 (1994).

\bibitem{Buba05} M. Buballa, Phys. Rep. \textbf{407}, 205 (2005).

\bibitem{Logo13} D. Logoteta, C. Provid\^{e}ncia, and I. Vida\~{n}a,
Phys. Rev. C \textbf{88}, 055802 (2013).

\bibitem{Bona12} L. Bonanno and A. Sedrakian, Astron. Astrophys. \textbf{539}, A16 (2012).

\bibitem{Fuku08} K. Fukushima, Phys. Rev. D \textbf{77}, 114028 (2008); \textbf{78}, 039902(E) (2008).

\bibitem{Ueda13} H. Ueda, T. Z. Nakano, A. Ohnishi, M. Ruggieri, and K. Sumiyoshi,
Phys. Rev. D \textbf{88}, 074006 (2013).

\bibitem{Buba15} M. Buballa and S. Carignano, Prog. Part. Nucl. Phys. \textbf{81}, 39 (2015).

\bibitem{Pagl08} G. Pagliara and J. Schaffner-Bielich, Phys. Rev. D \textbf{77}, 063004 (2008).

\bibitem{Abuk09} H. Abuki, R. Gatto, and M. Ruggieri, Phys. Rev. D \textbf{80}, 074019 (2009).

\bibitem{Masu13} K. Masuda, T. Hatsuda, and T. Takatsuka, Astrophys. J. \textbf{764}, 12 (2013);
Prog. Theor. Exp. Phys. \textbf{2013}, 073D01 (2013).

\bibitem{Mene14} D. P. Menezes, M. B. Pinto, L. B. Castro, P. Costa, and C. Provid\^{e}ncia,
Phys. Rev. C \textbf{89}, 055207 (2014).

\bibitem{Hell14} T. Hell and W. Weise, Phys. Rev. C \textbf{90}, 045801 (2014).

\bibitem{Chu15} P. C. Chu, X. Wang, L. W. Chen, and M. Huang,
Phys. Rev. D \textbf{91}, 023003 (2015).

\bibitem{Pere16} R. C. Pereira, P. Costa, and C. Provid\^{e}ncia,
Phys. Rev. D \textbf{94}, 094001 (2016).

\bibitem{Suga94} Y. Sugahara and H. Toki,
Nucl. Phys. A \textbf{579}, 557 (1994).

\bibitem{Shen02} H. Shen, Phys. Rev. C \textbf{65}, 035802 (2002).

\bibitem{Shen11} H. Shen, H. Toki, K. Oyamatsu, and K. Sumiyoshi,
Astrophys. J. Suppl. \textbf{197}, 20 (2011).

\bibitem{Demo10} P. B. Demorest, T. Pennucci, S. M. Ranson, M. S. E. Roberts,
and J. W. T. Hessels, Nature (London) \textbf{467}, 1081 (2010).

\bibitem{Fons16} E. Fonseca, T. T. Pennucci, J. A. Ellis, I. H. Stairs, D. J. Nice,
 S. M. Ransom, P. B. Demorest, Z. Arzoumanian, K. Crowter, T. Dolch,
 R. D. Ferdman, M. E. Gonzalez, G. Jones, M. L. Jones, M. T. Lam,
 L. Levin, M. A. McLaughlin, K. Stovall, J. K. Swiggum, and W. Zhu,
 Astrophys. J. \textbf{832}, 167 (2016).

\bibitem{Anto13} J. Antoniadis, P. C. C. Freire, N. Wex, T. M. Tauris, R. S. Lynch,
 M. H. van Kerkwijk, M. Kramer, C. Bassa, V. S. Dhillon, T. Driebe,
 J. W. T. Hessels, V. M. Kaspi, V. I. Kondratiev, N. Langer, T. R. Marsh,
 M. A. McLaughlin, T. T. Pennucci, S. M. Ransom, I. H. Stairs, J. van Leeuwen,
 J. P. W. Verbiest, and D. G. Whelan,
 Science \textbf{340}, 1233232 (2013).

\bibitem{Weis12} S. Weissenborn, D. Chatterjee, and J. Schaffner-Bielich,
Nucl. Phys. A \textbf{881}, 62 (2012).

\bibitem{Rehb96} P. Rehberg, S. P. Klevansky, and J. H\"{u}fner,
Phys. Rev. C \textbf{53}, 410 (1996).

\bibitem{Baym71} G. Baym, H. A. Bethe, and C. J. Pethick,
Nucl. Phys. A \textbf{175}, 225 (1971).

\end{thebibliography}
\end{document}